\documentclass[%
 reprint,
 amsmath,amssymb,
 aps,
]{revtex4-2}

\usepackage{graphicx}% Include figure files
\usepackage{dcolumn}% Align table columns on decimal point
\usepackage{bm}% bold math
\usepackage{float}
\usepackage{xcolor}
\usepackage{soul} 
\usepackage{siunitx}

\begin{document}

\title{Stoichiometry Dependent Properties of Cerium Hydride: An Active Learning Developed Interatomic Potential Study}

\author{Brenden W. Hamilton}
 \email{brenden@lanl.gov}
 \author{Travis E. Jones}
\author{Timothy C. Germann}
\author{Benjamin T. Nebgen}
 \affiliation{Theoretical Division, Los Alamos National Laboratory, Los Alamos, New Mexico 87545, USA}

\date{\today}

\begin{abstract}

Cerium hydride has a variety of interesting properties, including a known lattice contraction and densification with increasing hydrogen content.
However, precise stoichiometric control is not experimentally straightforward and {\it ab initio} approaches are not computationally feasible for many properties such as melting and low temperature diffusion.
Therefore, we develop a machine-learned interatomic potential for cerium hydride that is valid for H to Ce ratios from 2.0 to 3.0.
A query-by-committee active learning approach is used to develop the training set.
Leveraging classical molecular dynamics simulations, we assess a range of properties and provide fundamental mechanisms for the trends with stoichiometry.
A majority of the properties follow the trend of lattice contraction, being governed by the stronger lattice binding induced by adding octahedral atoms.

\end{abstract}

\maketitle

\section{Introduction}

Since their inception in the 1950s\cite{Alder1957Phase}, atomistic simulations such as molecular dynamics (MD), have become a broadly applicable tool for providing crucial insight into physical and chemical processes governed at the nanoscale\cite{hollingsworth2018molecular,van1990computer,karplus1990molecular,perez2009accelerated,wolf2005deformation,Hamilton2021Review,hamilton2024high}.
The predictive power of MD, in general, depends on the quality of the interatomic potential, which governs the forces and energies of the atoms, and hence, the overall dynamics. 
Classical interatomic potentials, which have defined functional forms, have made great scientific advances\cite{stoneham1986interatomic,muser2023interatomic}, but are limited in terms of broadness and flexibility.

Machine learning (ML) has recently appeared as a well established alternative model framework for physical simulations\cite{behler2007generalized,carleo2019machine,mishra2025unveiling,karniadakis2021physics,hamilton2023using,lafourcade2023multiscale}. 
ML interatomic potentials (MLIAPs) require significant data generation with high level quantum chemistry approaches for training, and are often more expensive than their classical counterparts, 
but the lack of a functional form for neural network type models
allows for a significant increase in accuracy, reaching that of {\it ab initio} methods in a classical framework\cite{fedik2022extending,deringer2019machine,zuo2020performance,f2025improving,hamilton2023high}.

The key in developing robust and useful MLIAPs is in the data generation and sampling process.
The chosen structures in the training set need to span the relevant region of the potential energy surface, such that all possible configurations of the atoms in question are an interpolation.
For simple metallic systems, this often involves several crystal structures at a range of pressures and temperatures, as well as liquid and defect structures.
Designing these training sets by hand can be a daunting task.
Physics-informed or data-driven generation algorithms provide an automated and robust route to training set design, using the MLIAP itself to drive discovery of phase space\cite{kulichenko2024data}.
Active learning (AL) type methods attempt to generate data in areas in which the existing model has high uncertainty, adding the data that is most likely to improve model robustness.
By assessing the underlying uncertainty, AL-driven sampling can limit the amount of expensive labeling needed in order to optimize a MLIAP workflow\cite{zhang2019active}.
AL routines\cite{druck2009active} with physics-driven sampling such as query-by-committee\cite{seung1992query,smith2018less,smith2021automated}, maximum entropy\cite{montes2022training,karabin2020entropy}, 
and genetic algorithms\cite{browning2017genetic,Chan2019Machine} are commonly used to build highly accurate MLIAPs.

Despite the increase in high-level tools for developing MLIAPs, materials with complex structures or phenomena still pose a challenge in developing a model that spans all of the system.
Cerium and cerium hydride are one such example, as cerium undergoes a variety of complex phase transformations\cite{casadei2012density,cadien2013first,lanata2013gamma,amadon2006alpha,hamilton2026Mott},
and cerium hydride is stable for a wide variety of stoichiometry\cite{sarussi1993kinetics,gurel2009volume,chen2021high}.
CeH$_X$ is stable for a range of $X$ from roughly 1.95-3.00 in the FCC phase\cite{lundin1966thermodynamics} whereas CeH$_2$ exists in a fluorite structure\cite{holleycrystal}.
CeH is also known to transition from a metal to a semiconductor at CeH$_{2.75}$\cite{libowitz1972temperature}, and has a stable BCC hydride phase at temperatures above 1000 K and for $X$ less than 2.0\cite{manchester1997h}.
There is also a known lattice contraction with increasing hydrogen content when the metal atoms are in the FCC configuration\cite{ao2012lattice}.
The large size of the octahedral sites compared to the hydrogen ions, as well as participation of the Ce 5d electron in bonding, leads to a shrinking of the lattice to form chemical bonds.
This has also been predicted to have a stiffening effect, increasing the elastic constants\cite{gurel2009volume}.

Despite a wide range of interesting physics, experimentally controlling the hydride stoichiometry and working with the material poses a difficult task,
leading to very sparse data on how changing the stoichiometry can vary CeH$_X$ properties.
Here, we utilize active learning to generate an MLIAP training set for cerium hydrides at the PBE+$U$ level of DFT, focusing mainly on a wide range of temperatures and densities for the FCC hydrides
at a range of stoichiometries. We then utilize the potential to assess a wide range of properties and their stoichiometric dependence for an unprecedented extent of stoichiometries,
assessing the underlying mechanisms for various trends.

\section{Methods and Interatomic Potential}

\begin{figure*}[htbp]
  \includegraphics[width=0.9\textwidth]{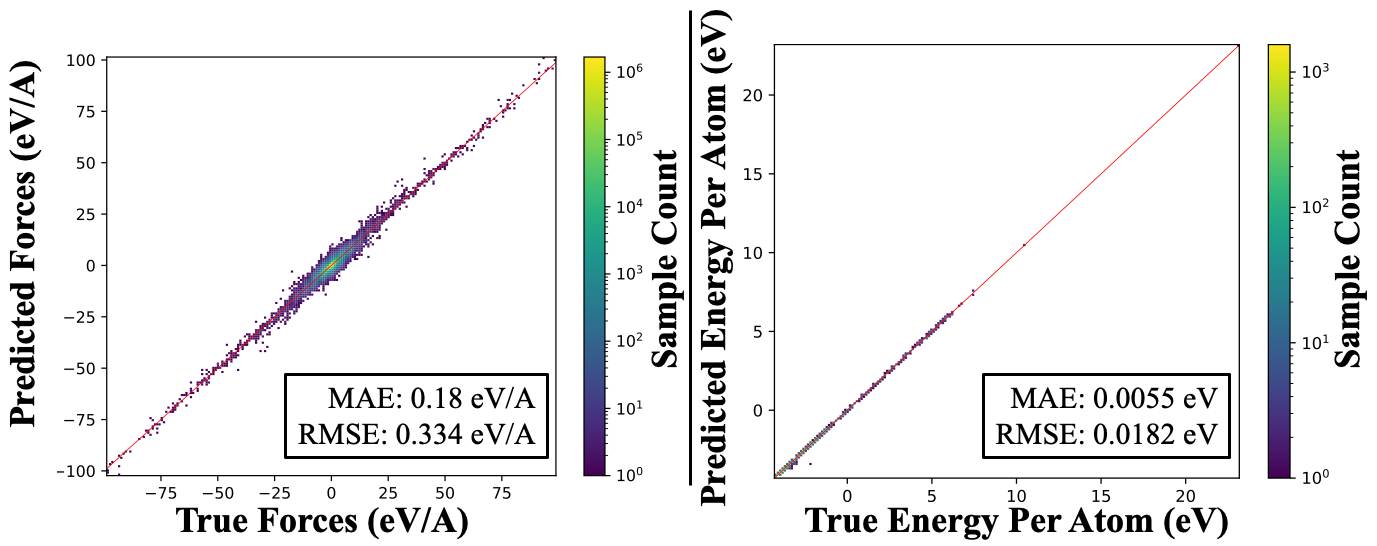}
  \caption{Force and energy parity histograms for the HIPNN CeH model.}
  \label{fig:Fg1}
\end{figure*}

We utilize an query by committee-based AL approach\cite{smith2018less} to train a hierarchically interacting particle neural network (HIPNN)\cite{lubbers2018hierarchical} model for the Ce/CeH$_X$ system.
HIPNN utilizes an end-to-end, graph-based, message-passing neural network approach.
We employ eight HIPNN models training with different, random test-train splits to sample uncertainty in the model during active learning. 

The HIPNN model was trained using both energy and force terms.
During training, we employ early stopping to prevent overfitting and anneal the learning rate to ensure a high-fidelity fit.
The learning rate is decreased by a factor of 0.5 with a patience of 15 epochs.
The model training is considered converged when the loss ceases to drop with a termination patience of 50 epochs.
Our HIPNN network structure consists of 2 interaction layers and 3 atom layers with 24 sensitivity functions. A 5.5 \AA\ cutoff is used.
Structures are filtered out of the data set if they contain atomic distances less than 0.35 \AA\ or have a maximum force above 100 eV/\AA.

To build the initial training, the AL approach samples Ce crystals in FCC, BCC, and dHCP structures and CeH$_X$ structures with varying stoichiometry.
All cells are $3\times3\times3$ replications of the unit cell. 
Lattice parameters are fractionally re-mapped to range from 0.5 to 2.5 times the minimum energy lattice parameter on the cold curve for each phase.
Deviatoric deformations are randomly applied to individual lattice directions, chosen from a normal distribution with a standard deviation of 0.25 \AA.
Atoms are individually displaced from the perfect crystal lattice sites with random displacements, following a normal distribution with a standard deviation of 0.2 \AA.
An initial model is trained on 500 randomly selected structures before AL begins.

Using the initial model (and each subsequent, most recent model), classical MD simulations are spawned the same way as the DFT runs by selecting a random structure and displacing the atoms.
The initial temperature is randomly chosen to be between 200 K and 500 K. Simulations are set to run for 100 ps with temperatures ramping to a final value between 2500 K and 4500 K (randomly chosen following a uniform distribution).
Density is also altered throughout the simulation to end between 0.5 g/cm$^3$ and 18 g/cm$^3$.
During each step, the forces and energy are calculated using all eight of the HIPNN models trained in the most recent training.
If at any point, the standard deviation of a force or the energy is greater than 0.2 eV/\AA\ and 0.5 eV, respectively, the MD simulation ends and the uncertain structure is used as the atomic coordinates for a DFT calculation to label new training data.
Every 100 uncertain samples, the DFT energy and forces are calculated and added to the training data, a new set of models are trained, and MD is then run with this newer model.
This cycle is continued until MD can be successfully run across the temperature and density space with a minimal number of uncertain structures being found. 

All structures flagged by the AL approach are labeled for forces and energy with DFT.
We perform DFT calculations at the PBE+$U$ level\cite{Perdew1996GGA, anisimov1991band} using VASP\cite{Kresse1996PlaneWave}.
Calculations are performed on a standard {\bf k}-point mesh grid spacing of 0.2 \AA$^{-1}$ and a 1000 eV plane-wave cutoff.
Gaussian smearing was used with a smearing width of 0.20 eV.
The Hubbard $U$ parameter was set to 4.0 eV.
Convergence criteria was set to a cutoff of \num{1e-6} eV.
All other parameters were kept at the default values for conducting PBE+$U$ calculations in VASP.
The calculations were conducted without magnetic moments, which prevents DFT from capturing the $\gamma$ to $\alpha$ phase transition in the metal\cite{tran2014nonmagnetic}.
However, initial comparisons for the hydride systems at ambient to moderate pressures showed little to no effect on forces and energies from adding magnetic moments.

The final model, in which error histograms are presented in Figure 1, consists of 18,068 structures, where 10\% was used for testing and validation each.
The force and energy MAE values are 0.18 eV/\AA\ and 0.0055 eV, respectively.

For MD simulations, all systems are a $10\times10\times10$ replication of the unit cell and a 0.2 fs timestep is used, except for melting simulations, in which the structure is a $15\times15\times30$ replication of the unit cell.
Initial structures are equilibrated at 300 K under isochoric conditions for 10 ps, followed by 10 ps of isobaric conditions at 1.0 bar.

Elastic constants are calculated by deforming the cell along deformation paths related to the six Voigt deformation components. Each deformation is instantaneous to a 0.5\% strain and held at
a constant volume, constant temperature state at 300 K for 100 ps. The last 50 ps is used to average the stress response and calculate the elastic constant.

Melting temperatures are measured using the two-phase method, in which half the system is melted at 2500 K and the other half held static over 50 ps to create an interface.
The system is then thermalized for 10 ps at an initial guess at the melting temperature, then allowed to relax under adiabatic conditions until equilibrium conditions are met.
The cell boundaries are allowed to relax to maintain 1.0 bar of pressure.
Diffusion properties are calculated by tracking the mean squared displacement of every hydrogen atom over 500 ps at isothermal-isobaric conditions, ignoring the first 50 ps of the trajectory in which the cell volume and temperature are equilibrating.

\section{Mechanical Properties}

The simplest of the CeH stoichiometry-dependent properties to assess is the lattice contraction with increasing octahedral hydrogen.
As additional hydrogen atoms are included in the lattice, the density is known to increase due to a lattice parameter contraction\cite{cheetham1972neutron,zamir1984proton}.
As small hydrogen atoms are placed in the large Ce lattice octahedral sites and bond with the Ce 5d electrons, the bonding favors a reduction in the atomic distances\cite{ao2012lattice}.
Figure 2 shows our density and cubic lattice parameter predictions at 300 K for a wide range of cerium hydride stoichiometries, showing that we capture these trends.
The increased resolution in stoichiometric space than is accessible through the scalability of the MD potential allows us to more finely assess these
trends of density, which is not linear with increasing hydrogen, but sub-linear, tapering off as the ratio approaches 3 hydrogen atoms per cerium (12 per unit cell).
As expected from the density, the cell lattice constant also contracts with increasing hydrogen in a non-linear fashion.
While the exact lattice parameter values are off from experimental values by roughly 0.1 \AA, the contraction of 0.04 \AA\ from CeH$_2$ to CeH$_3$ agrees very well with experiments\cite{korst1966rare}.

One physical explanation for this tapering effect at higher hydrogen content is that increasing the number of bonds between octahedral hydrogen and Ce atoms has diminishing returns
due to the overlapping strain fields.
In the cases of a single octahedral atom placed in bulk CeH$_2$, there will be a long range strain field as the lattice adjusts to compensate for the local contraction around the octahedral hydrogen.
For small amounts of octahedral occupation, the strain fields will have minimal interaction, leading to maximal lattice contraction.
As the octahedral sites are nearly full, additional hydrogen atoms will result in additional strain fields that are almost entirely canceled out by neighboring octahedral hydrogen, leading to almost no lattice contraction.
These non-linear trends in density/lattice constant will parallel several other properties, as we discuss below.

\begin{figure}[htbp]
  \includegraphics[width=0.45\textwidth]{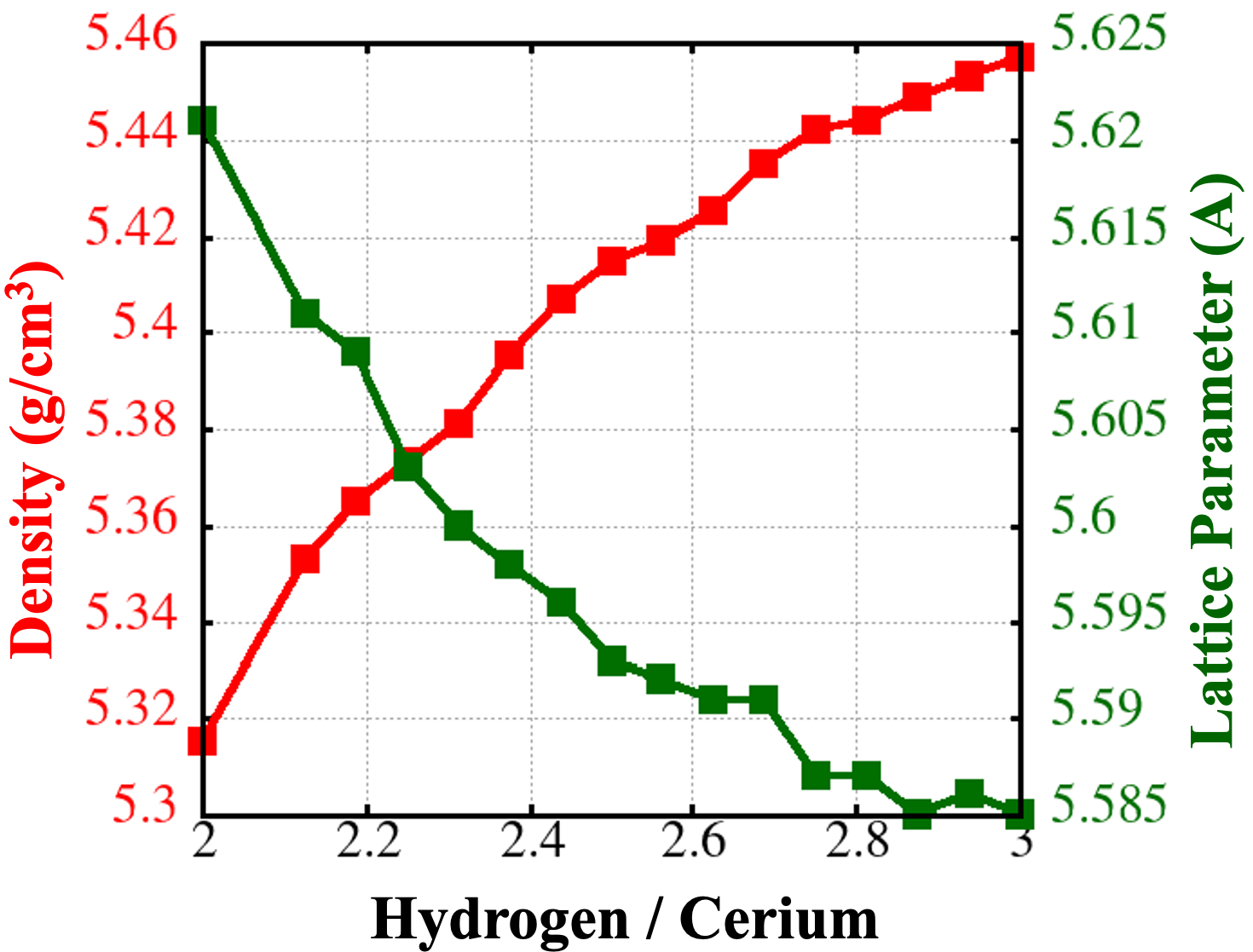}
  \caption{Equilibrium, 300 K density and lattice parameters for a range of CeH$_X$ stoichiometries.}
  \label{fig:Fg2}
\end{figure}

For the 300 K elastic constants, as shown in Figure 3, we see a range of trends for $C_{11}$, $C_{12}$, and $C_{44}$.
For the $C_{11}$ elastic constant, we see a strong rise with increasing hydrogen from CeH$_2$ to CeH$_{2.5}$, and then a local maximum with a slight drop in $C_{11}$ nearing CeH$_3$.
For $C_{12}$, or the resistance to lateral contraction with uniaxial deformation, there is a direct, linear increase with increasing hydrogen content.
As hydrogen is added to octahedral locations, the lattice contracts and is more tightly bonded, leading to a more rigid lattice that builds more transverse stresses to a uniaxial deformation.
For $C_{44}$, we see a similar tapering off in the increase with hydrogen content as we do with density.
$C_{44}$, the material's resistance the shear deformation, follows the effect of the lattice contractions, signifying that they are both directly governed by the local strain fields induced by the octahedral hydrogen,
where additional Ce-H bonds increase the rigidity and prevent shape deformation.

\begin{figure}[htbp]
  \includegraphics[width=0.45\textwidth]{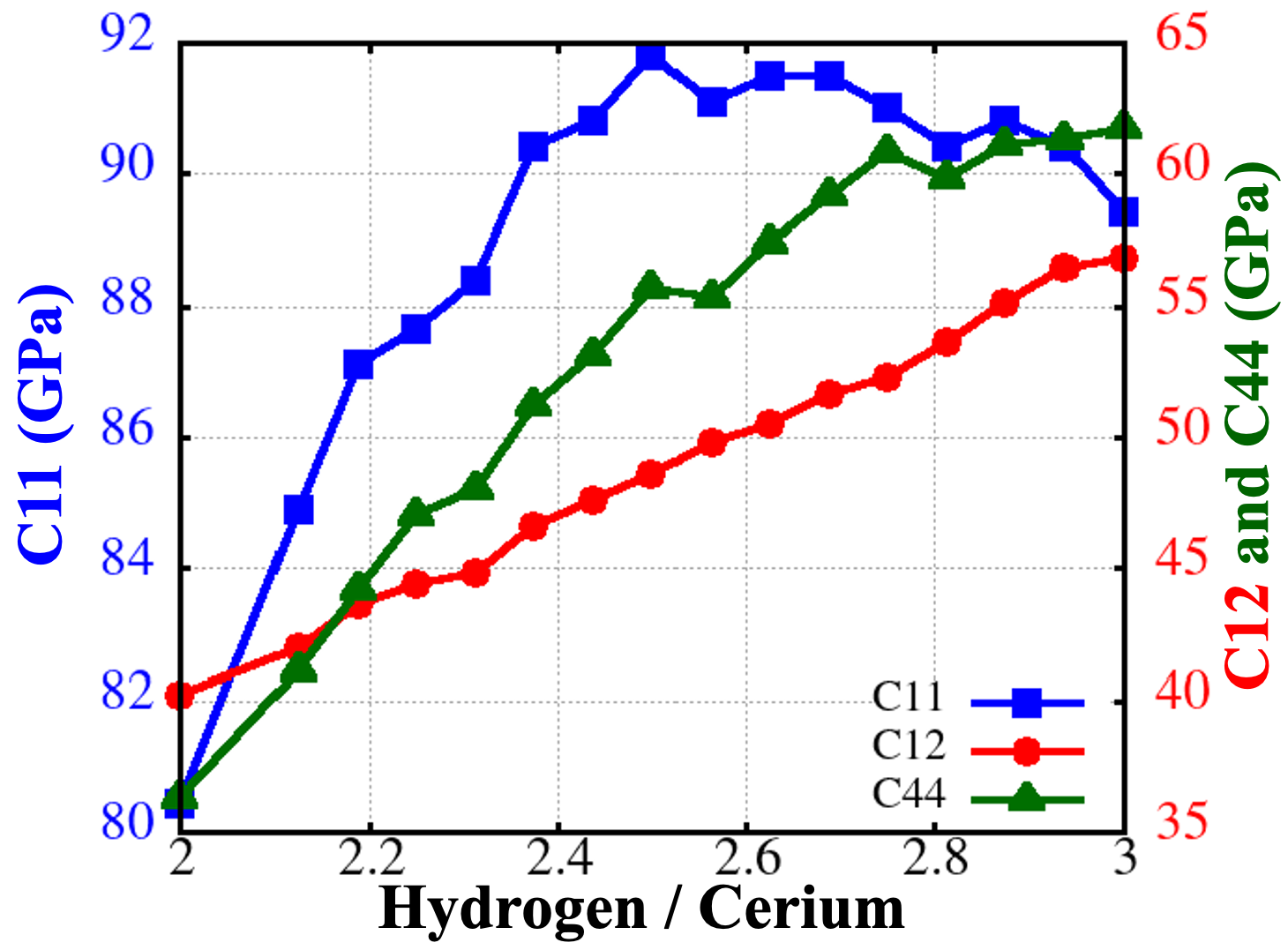}
  \caption{Equilibrium, 300 K elastic constants for a range of CeH stoichiometries. $C_{11}$ corresponds to the left $y$ axis, $C_{12}$ and $C_{44}$ to the right axis.}
  \label{fig:Fg3}
\end{figure}

Lastly, for mechanical properties, we sample the pressure response over stoichiometry for a wide range of volumes.
Figure 4 shows that there is very little deviation in the $P-V$ response across stoichiometry for low pressures.
It is not until large compressions and pressures reaching above 30 GPa that noticeable deviation occurs.
Higher hydrogen concentrations quickly reach higher pressures for a given compression.
The additional hydrogen occupation increases the material stiffness, which is not overly apparent for low compression,
but leads to a much higher resistance to compression as high pressures.

\begin{figure}[htbp]
  \includegraphics[width=0.45\textwidth]{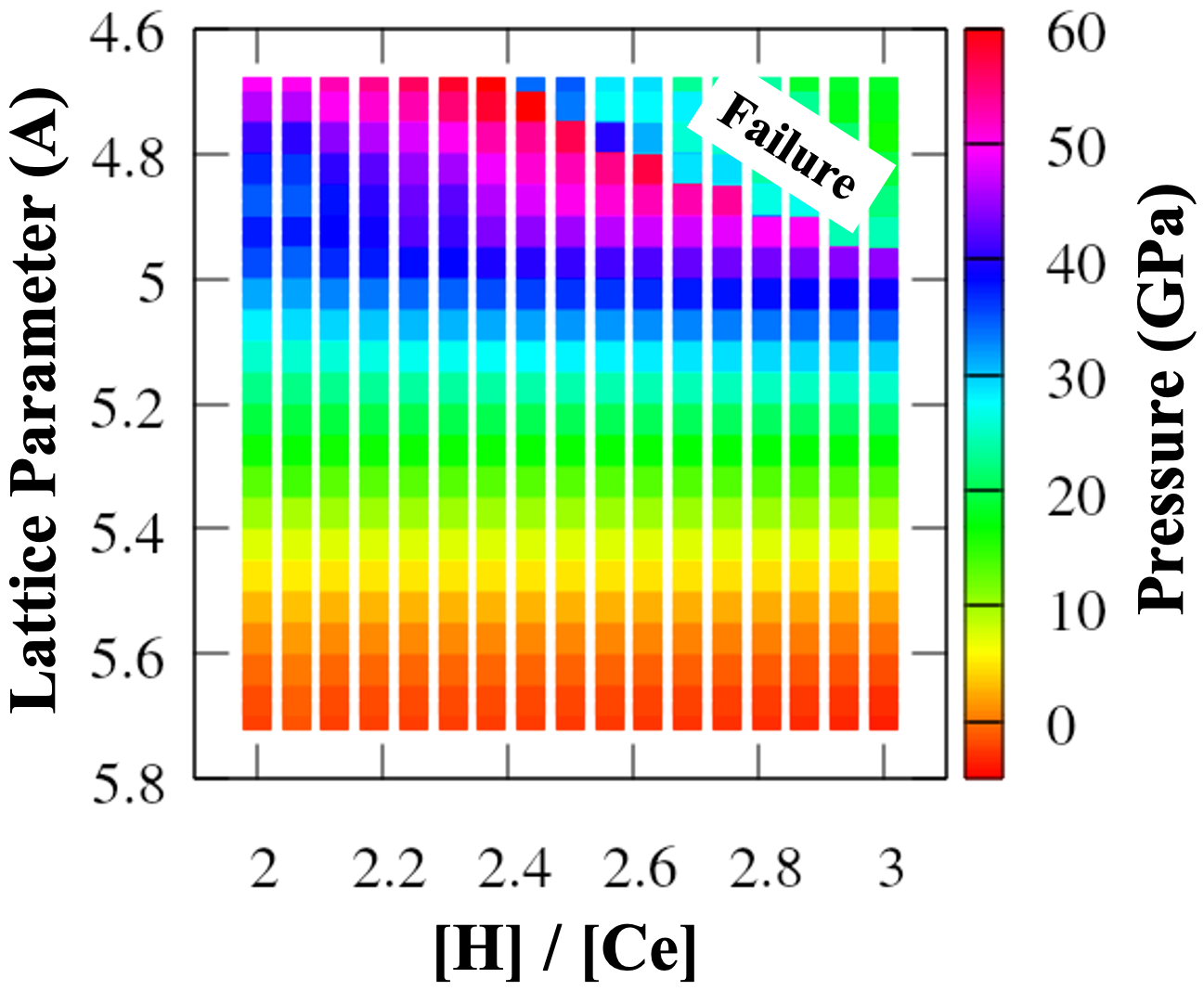}
  \caption{Heat map of the pressure response of CeH$_X$ for varying stoichiometry and hydrostatic compressions. Points in the upper right corner undergo compression-induced yielding.}
  \label{fig:Fg4}
\end{figure}

\section{Melting and Hydrogen Diffusion}

In theory, numerous, large DFT-MD calculations could have been used to assess the properties shown in the previous sections, and the MLIAP used here only greatly lowered the computational cost.
However, for the remainder of this manuscript, we turn to properties that require much large length and time scales to assess,
in which the scalability of the potential can be leveraged.

Assessing the melting temperature requires simulating large, non-equilibrium systems
for upwards of nanosecond timescales.
Figure 5 shows the predicted melting temperatures of CeH$_X$ as a function of stoichiometry.
We see a strong initial rise in melting temperature, followed by a tapering off at higher amounts of hydrogen.
This, similar to other properties studied here, follows the trend of density/lattice contraction.
As more octahedral hydrogen are added to the system, the lattice is more strongly bonded, leading to the lattice contraction, as well as a need for more thermal energy to overcome the cohesive energy of the system and induce melting.

In the case of melting, stoichiometries are not assessed above 2.7 H per Ce.
In general, liquid Ce and H$_2$ gas will want to phase separate.
At lower hydrogen concentrations, the mixed liquid phase is metastable enough to assess a melting point.
At high hydrogen concentrations, the liquid begins to phase separate on the timescales of the two-phase melting simulations.
This leads to two-phase melting simulations where the liquid phase is not the same material as the solid hydride, preventing prediction of melting.

\begin{figure}[htbp]
  \includegraphics[width=0.45\textwidth]{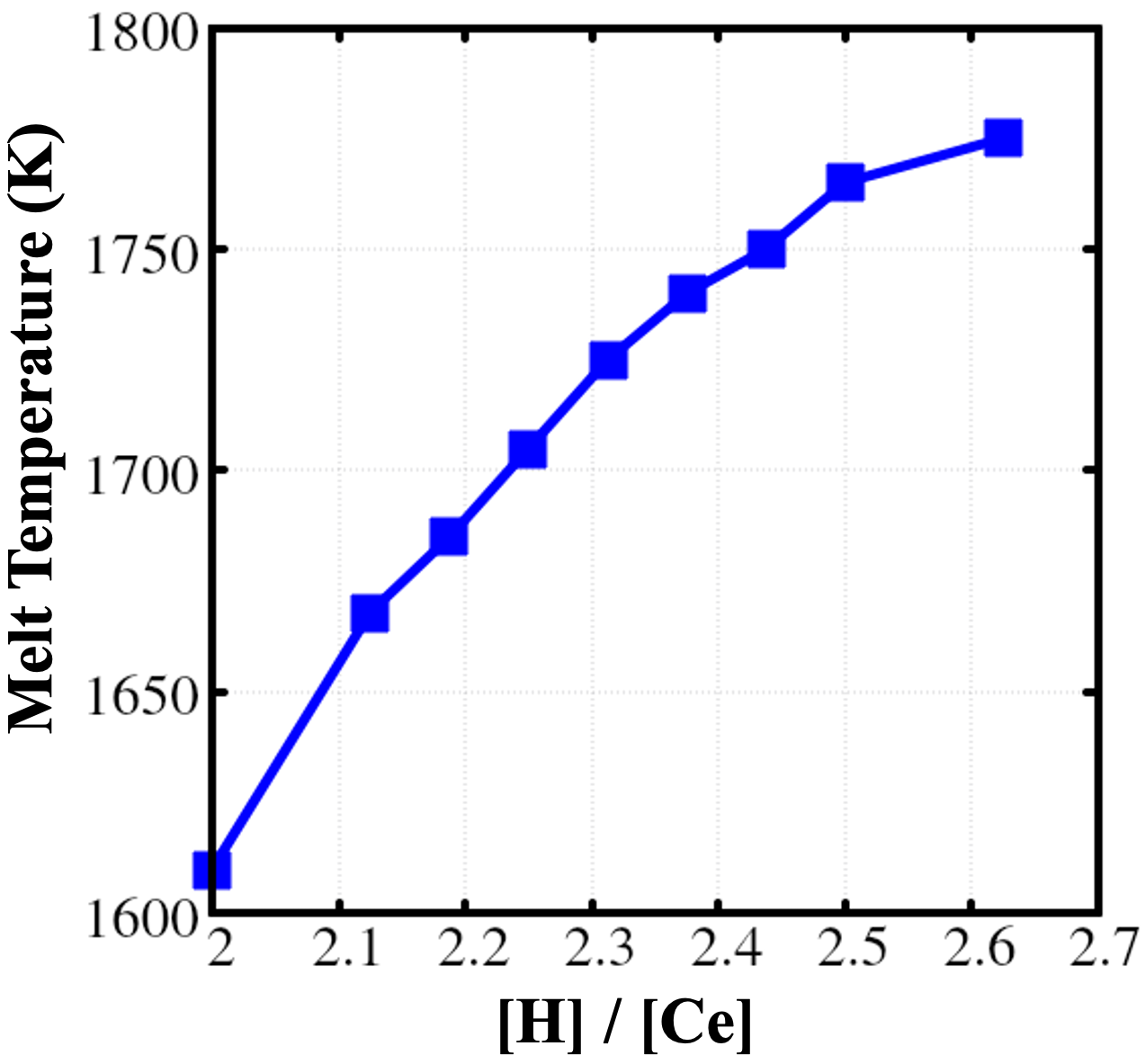}
  \caption{Stoichiometric dependent melting points for CeH$_X$ at 1 bar of pressure.}
  \label{fig:Fg5}
\end{figure}

\begin{figure}[htbp]
  \includegraphics[width=0.45\textwidth]{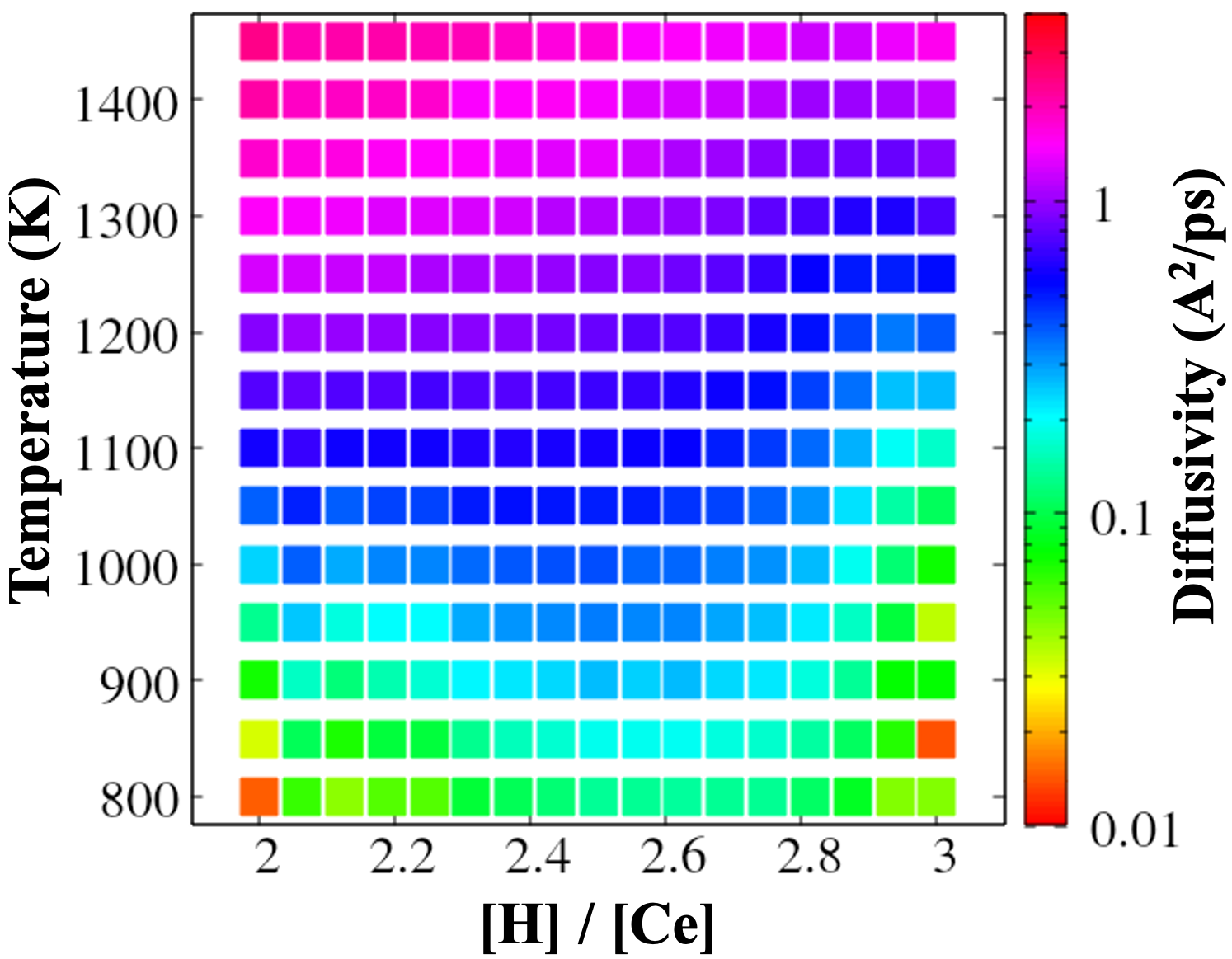}
  \caption{Heat map of the diffusivity of CeH$_X$ for varying stoichiometry and temperature.}
  \label{fig:Fg6}
\end{figure}

Turning to hydrogen diffusion, Figure~\ref{fig:Fg6} shows the calculated diffusivity for a variety of stoichiometries as a function of temperature.
Interestingly, we find distinct low and high temperature regimes.
At high temperatures, CeH$_2$ has the highest diffusivity as the thermal energy is sufficient to drive hydrogen hopping out of tetrahedral holes.
Since all four octahedral locations are vacant, the tetrahedral hydrogen have active channels in which to diffuse through, resulting in the most un-impeded motion of the stoichiometries studied here, 
as $D = D_{eff} * C_v$, where $D_{eff}$ is constant at a single temperature and $C_v$ is the vacancy concentration where unoccupied octahedral sites are considered vacancies of CeH$_3$.
As octahedral hydrogen are added, this restricts the physical space for diffusion, in accordance with classical lattice diffusion.

Conversely, at low temperatures, the is a maximum in diffusivity at the intermediate stoichiometries, between 2.5 and 2.6 H per Ce, with extremely low diffusivity in CeH$_2$.
Energetically, hydrogen atoms are more stable in the Ce tetrahedral sites, leading to the octahedral $\rightarrow$ tetrahedral having a lower activation energy than tetrahedral $\rightarrow$ octahedral diffusion paths.
Thus, in CeH$_2$, at low temperatures, all hydrogen are not well activated. As octahedral hydrogen are added, they are more readily thermally activated to move locations.
The two main possibilities for the non-monotonic behavior at low temperature are then a concerted mechanism between a tetrahedral and octahedral hydrogen or that the octahedral hydrogen occupancy inherently lowers the energy barrier for tetrahedral diffusion through
the local displacement of phonons from the octahedral occupation-induced strain field.

Qualitatively, the concerted diffusion motion would consist of the octahedral hydrogen pushing the tetrahedral hydrogen into an open octahedral site and then moving to the tetrahedral location.
This mechanism keeps the tetrahedral locations occupied, maintaining the low energy configuration and keeping the local crystal symmetry the same.
For the lower energy barrier mechanism, adding octahedral hydrogen could physically alter the space in which tetrahedral hydrogen have to move.
The octahedral hydrogen atoms bond strongly with the cerium atoms, contracting the lattice. This could drive cerium atoms towards the octahedral atoms, weakening its hold on the tetrahedral atoms and giving a larger volume to move.
In both cases, initially adding more octahedral hydrogen enables more total diffusion events up until the octahedral locations are half full.
Since both mechanisms require the tetrahedral hydrogen to be adjacent to both an occupied and unoccupied octahedral location, adding octahedral atoms passed half occupancy will remove available diffusion channels (via removal of vacancies)
and lower the overall diffusivity, leading to the maxima shown in Figure~\ref{fig:Fg6}.

\section{Conclusions}

Here, we present a machine learned interatomic potential for cerium hydride, using it to assess a variety of material properties as a function of stoichiometry.
We find that the vast majority of these properties are governed by the lattice contraction and stronger binding from the inclusion of more octahedral hydrogen.
Elastic constants, stiffness, and melting all directly follow the same trends as the lattice contraction, which is a fast growing response at low octahedral occupation,
but tapers off and grows slower after half occupation.
Diffusion is the one property that does not follow these trends. It shows an opposite trend with increasing octahedral occupation at high temperatures causing a drop in diffusivity in accordance with classical diffusion, while at low temperatures
a local maximum in diffusivity around half occupation of the octahedral holes is observed, demonstrating the hydrogen occupation drives a change in activation energy for diffusion.

\section{Acknowledgments}

Funding for this project was provided by the Advanced Simulation and Computing Physics and Engineering Models project (ASC-PEM). Partial funding was provided by ASC Computational Systems \& Software Environment (CSSE).
This research used resources provided by the Los Alamos National Laboratory Institutional Computing Program, which is supported by the U.S. Department of Energy National Nuclear Security Administration under Contract No. 89233218CNA000001.
Approved for Unlimited Release LA-UR-26-21169

\bibliography{references}

\end{document}